# Activity of (6478) Gault during January 13 – March 28, 2019

Oleksandra Ivanova[a,b,c,*], Yuri Skorov[d], Igor Luk'yanyk[c], Dušan Tomko[a], Marek Husárik[a], Jürgen Blum[d], Oleg Egorov[e], Olga Voziakova[e]

[a] *Astronomical Institute of the Slovak Academy of Sciences, SK-05960 Tatranská Lomnica, Slovak Republic*

[b] *Main Astronomical Observatory of the National Academy of Sciences of Ukraine, 27 Zabolotnoho Str., 03143 Kyiv, Ukraine*

[c] *Astronomical Observatory of Taras Shevchenko National University of Kyiv, 3 Observatorna Str., 04053 Kyiv, Ukraine*

[d-] *Technische Universität Braunschweig, Institute for Geophysics and Extraterrestrial Physics, Mendelssohnstr. 3, D-38106 Braunschweig, Germany*

[e-] *Sternberg Astronomical Institute, Lomonosov Moscow State University, Universitetsky pr. 13, Moscow 119234, Russia*

[*] Corresponding Author. E-mail address: oivanova@ta3.sk

Pages: 30
Tables: 3
Figures: 10
**Proposed Running Head:** Activity of (6478) Gault

**Editorial correspondence to:**

Dr. Oleksandra Ivanova

Astronomical Institute of the Slovak Academy of Sciences,

SK-05960 Tatranská Lomnica, Slovak Republic

E-mail address: oivanova@ta3.sk

**Highlights**

- Different observations of active asteroid (6478) Gault were performed in the optical range with telescopes of 0.61 m and 2.5 m diameter in imaging mode between January 15 and March 28, 2019.

- Two distinct comet-like dust tails were detected during the observations. A similar morphology was observed in previous active phase of the asteroid in 2013 and 2016. The morphology of (6478) Gault is typical for most of the active asteroids (133P, 238P, 311P): it is a point-like nucleus with a long thin dust tail.

- Low measured values of $Af\rho$ (from 32 to 46 cm) indicate weak activity compared to typical dust production for Jupiter Family Comets at a similar heliocentric distance.

- The asteroid is redder than the sun. The colour indices B-V, V-R and B-R fall down significantly with increasing distance from the asteroid in the region around ~10 000 km.

- The dynamical evolution of Gault's orbit reveals relatively minor changes during the last 100 millennia.

- Irregular activity may be caused by meteoroid impacts on the surface of Gault.

# ABSTRACT

We present the results of photometric observations of active asteroid (6478) Gault performed at heliocentric distances from 2.46 to 2.30 au and geocentric distances from 1.79 to 1.42 au between January 15 and March 28, 2019. Observations were carried out at the 2.5-m telescope of SAI MSU (CMO) on January 15, 2019 and at the 1.3-m and 0.61-m telescopes (SPb) on February 6 and March 28, 2019, respectively. The direct images of the asteroid were obtained with the broad-band B, V and R filters. Comet-like structures were detected at all observation dates. Colour maps were built and colour variations along the tail for the observation made on January 15, 2019 were analyzed. The $Af\rho$ was calculated for the R filter, The evaluated value varies from 47 to 32 cm for the period from January to the end of March, 2019. The rotational period of the body is estimated from the light curve by different methods and is about 1.79 hr. Possible mechanisms of triggering Gault's activity are discussed.

# 1. Introduction

Asteroids, like comets, belong to the population of small bodies of the Solar system. For a long time, these objects were rather arbitrarily subdivided based on signs that have different nature and different validity. The most common classification (Davies et al., 1982; Tholen, 1989) takes into account a) the presence or absence of an extended gas/dust coma and/or tail(s), b) an abundance or lack of volatiles (ices), and c) orbit specificity (semi-major axis, eccentricity and inclination of the orbit). Now it is clear that these signs are not independent and often cannot be reliably determined. It is the so-called active asteroids (hereafter AA) that are recently discovered troublemakers: these objects have typical asteroid orbits, but show the activity of typical comets. Jewitt et. al (2015a) paid attention to this blurriness of the classification and suggested a simple two-parameter scheme that usefully describes the small-body populations. This empirical scheme is based on the Tisserand parameter and the presence of a coma. Following this idea, objects with asteroidal orbits and well-pronounced activity (i.e. mass loss) are attributed to active asteroids.

Note that for comets, a long-term activity caused by the sublimation of ices and the concomitant removal of dust (Whipple 1951) is usually in the focus of study, although short-term activities for comets are also observed. Manifestations of asteroid activity are perhaps more diverse. In all identified cases, these objects, like comets, lose a noticeable amount of material. That is, they are active, sometimes short-lived (Neslusan et. al., 2016), sometimes repeated (133P/Elst-Pizarro, 6478 Gault). Several mechanisms that evoke activity are discussed in the asteroid community: ice sublimation and accompanying dust removal (e.g., Hsieh et. al., 2010 considered these processes for 133P/Elst-Pizarro); rotational breakup (e.g., suggested by Drahus et al., 2015 for P/2012 F5 (Gibbs)), impact and resulting dust ejecta (e.g., Neslusan et. al., 2016 applied this hypothesis to explain the activity of (596) Scheila, Bodewits et.. al., 2011; Ishiguro et. al., 2011; Jewitt et. al., 2011 also studied this mechanism of activity), thermal fracture (e.g., suggested by Jewitt & Li, 2010; Li & Jewitt, 2013; Hui & Li, 2016 for (3200) Phaethon) and rotational fission of contact binary asteroids (Scheeres, 2007). The determination of the mechanism of activity is often possible, based on an analysis of the specific features of the observations. For example, impact activity, as a rule, is characterized by a short time interval. The activity caused by the sublimation of ice can be related to the amount of energy absorbed, and therefore to the orbital position and so on. A simple enumeration shows that determining one or possibly several mechanisms of asteroid activity is a complex and important task.

MThe main-belt asteroid (6478) Gault (hereafter Gault) was discovered on 1988 May 12 by Carolyn and Eugene Shoemaker (https://www.minorplanetcenter.net/db_search/show_object?object_id=6478). Its orbital semimajor axis, eccentricity, perihelion distance and inclination are 2.305 au, 0.194, 1.859 au and 22.8 degrees, respectively. Knezevic & Milani (2003) classified it as a stony S-type asteroid and a member of the Phocaea family. Ye et al., (2019) showed that the colour of Gault was more similar to that of C-type asteroids than S-types. The asteroid Gault has also been dynamically linked with the low-albedo Tamara family (Kleyna et al., 2019), which resides in the Phocaea region (Novaković, 2017). Members of this family are characterized by the highest orbital inclinations among all the families of the inner asteroid belt. The Tisserand parameter of Gault is $T_J$ = 3.46. This value is significantly larger than the nominal dividing line ($T_J = 3$) that separates comets ($T_J < 3$) from asteroids ($T_J > 3$) (see, e.g., Kresak 1980). The recent activity of Gault was detected on January 5, 2019 (Smith and Denneau, 2019). An archive-data check revealed the activity of the asteroid in December 2018. Based on a thorough analysis of the archive images taken in 2013, 2016 and 2017, Chandler et al. (2019) concluded that Gault shows sustained activity since 2013. They also noted that this asteroid is unique: we do not know other members of this family that would be active so long. This unusual behavior caused great interest. Therefore, several articles presenting asteroid observations and their analyses were already published in 2019 (Chandler et al., 2019; Ferrín et al., 2019; Jewitt et al., 2019; Hui et al., 2019; Lee , 2019; Marsset et al., 2019; Moreno et al., 2019; Sanchez et al., 2019; Ye at al., 2019; Kleyna et al., 2019).

In this paper, we supplement a number of observations and present original data analyses. The rest of the paper is organized as follows. Details of observations and reduction procedures are described in Section 1. Analysis of the observed photometric data is presented in Section 2. A comparison of the original results with published data and their discussion is presented in Section 3.

## 2. Observations and data reduction

The observation of Gault was carried out with different telescopes. We obtained photometric data with the 2.5-m telescope of the Caucasian Mountain Observatory (CMO) of SAI MSU (Russia) and with the 1.3-m and the 0.61-m telescopes at the Skalnaté Pleso Observatory (Slovakia). The main technical parameters of the telescopes and CCD detectors are listed in Table 1. The observing log is presented in Table 2, listing the mid-cycle time; the heliocentric ($r$) and geocentric ($\Delta$) distances; the phase angle of the object ($\alpha$); the position angle of the extended Sun-asteroid radius vector ($\varphi$); the filter; the total exposure time during the night ($T_{exp}$); the number of cycles of observations obtained during the night ($N$); and the telescope used. The schematic view of the asteroid orbit along with the Earth's, Mars and Jupiter's orbits is shown in Fig.1.

Table 1. Equipment of the observations.

| Telescope | Diameter [m] | CCD | Pixel size [μm×μm] | Scale ["/pix] | Field of view [´×´] |
|---|---|---|---|---|---|
| CMO | 2.5 | NBI4K E2V CCD44-82 | 15×15 | 0.155 (bin 1×1) | 10×10 |
| SPb | 1.3 | FLI Proline 230 | 15×15 | 0.57 (bin 2×2) | 9.8×9.7 |
| TSPs | 0.61 | SBIG-ST-10XME | 9×9 | 1.07 (bin 2×2) | 19×13 |

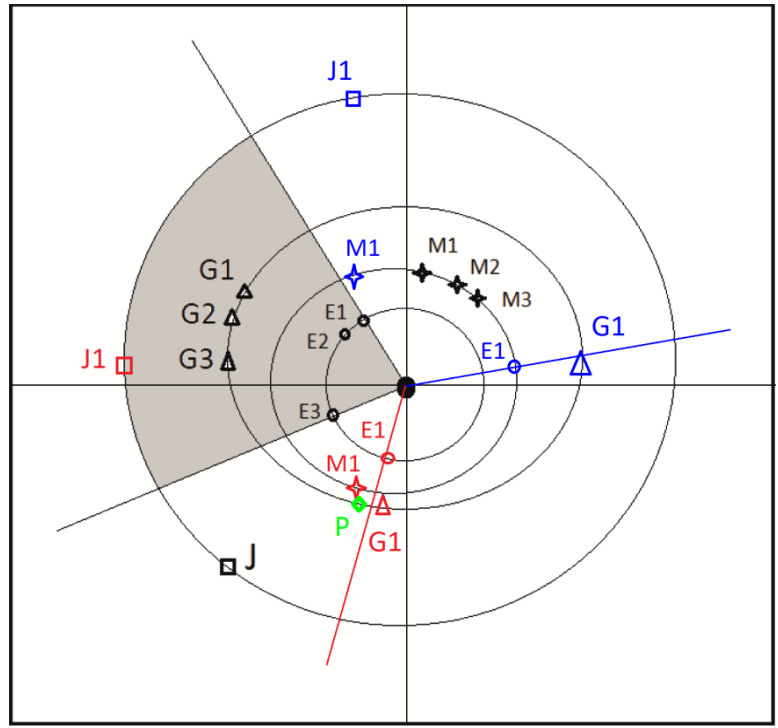

**Fig.1.** Schematic view of the asteroid orbit along with the Earth's (E), Mars (M) and Jupiter's (J), orbits. Positions of the asteroid and planets corresponding to the selected date of observations January 13, 2019 (1), February 6, 2019 (2) and March 28, 2019 (3) are marked in black. Corresponding positions for the previous activity observations on September 28, 2013 (blue marks) and June 10, 2016 (red marks) are also shown (Chandler et al., 2019). Green diamond (P) marks the perihelion of the asteroid orbit on January 3, 2020 ($q$=1.86 au).

Table 2. Log of the observations of asteroid (6478) Gault in 2019.

| Date, UT | r [au] | Δ [au] | α [deg] | φ [deg] | Filter | Texp [sec] | N | Instrument |
|---|---|---|---|---|---|---|---|---|
| 2019 Jan. 13 | 2.460 | 1.797 | 19.9 | 305.6 | B,V,R | 300 | 3, 3, 3 | 2.5-m CMO |
| 2019 Jan. 15 | 2.456 | 1.772 | 19.5 | 306.5 | B,V,R | 300 | 3, 3, 3 | 2.5-m CMO |
| 2019 Jan. 16 | 2.454 | 1.759 | 19.3 | 307.1 | B,V,R | 300 | 3, 3, 3 | 2.5-m CMO |
| 2019 Jan. 18 | 2.450 | 1.735 | 18.85 | 308.1 | B,V,R | 120 | 3, 3, 3 | 2.5-m CMO |
| 2019 Feb. 06 | 2.411 | 1.525 | 13.0 | 323.2 | V, R | 240 | 20, 20 | 1.3-m Pleso |
| 2019 Feb. 08 | 2.407 | 1.509 | 12.3 | 324.6 | B, V, R | 240 | 13, 13, 14 | 1.3-m Pleso |
| 2019 Mar. 23 | 2.316 | 1.405 | 12.9 | 90.34 | V, R | 240 | 6, 6 | 0.61-m Pleso |
| 2019 Mar. 27 | 2.307 | 1.421 | 14.2 | 93.63 | V, R | 240 | 10,10 | 1.3-m Pleso |
| 2019 Mar. 28 | 2.305 | 1.425 | 15.0 | 95.46 | R | 240 | 46 | 1.3-m Pleso |

We applied the reduction procedure - bias subtraction, dark-field correction (only for observation obtained at 0.61-m and 1.3-m telescopes), flat-field correction, and cleaning cosmic ray tracks in the standard manner, using IDL routines (e.g. Ivanova et al. 2016, 2017, 2019; Picazzio et al. 2019, Luk'yanyk et al. 2019). The morning sky was used to provide a flat-field correction for the non-uniform sensitivity of the CCD chip. The seeing value was measured as the average FWHM of several sample stars ranging from 1.3 to 3.5 arcsec during our observations, the data with a large seeing were not taken into account. The residual sky background was estimated with the use of an annular aperture. To perform an absolute flux calibration of the images, field stars were used. The stellar magnitudes of the standard stars were taken from the catalogue APASS (Henden et al., 2014). The photometric uncertainty of the catalogue depends on star brightness and is estimated to be from 0.01 to 0.2 mag on average. The images were binned in 2×2 pixels to improve the signal/noise (S/N) ratio of the measured signal. In January 2019, direct images of the asteroid were obtained with the broad-band Bessel B, V and R filters (http://lnfm1.sai.msu.ru/kgo/instruments/filters/KGO_FILTER_DATA.html). Johnson-Cousins B, V and R filters were used for observations obtained from February to March 2019.

## 3. Photometry

*Morphology*

R-band images of asteroid Gault are shown in Fig. 2. These data were obtained at the 2.5-m (CMO) telescope of SAI MSU on January 15, 2019 (A), at the 1.3-m (SPb) telescope on February 6 and March 28, 2019 (B and C), respectively. The resulting image is the sum of stacked particular images. The colour bar shows a relative intensity scale. The asteroid coma is compact and bright. On January 15, 2019, the asteroid presents only one tail T1. The asteroid exhibits a second faint tail (T2) on images obtained from the observations carried out on February 6, 2019, and on March 23, 2019. For observations made in January and February 2019, the dust tail (T1) is elongated in the anti-sunward direction, as can be seen in panels A and B. On March 28, 2019 (panel C), both tails (T1 and T2) change the orientation relative to the sun dramatically.

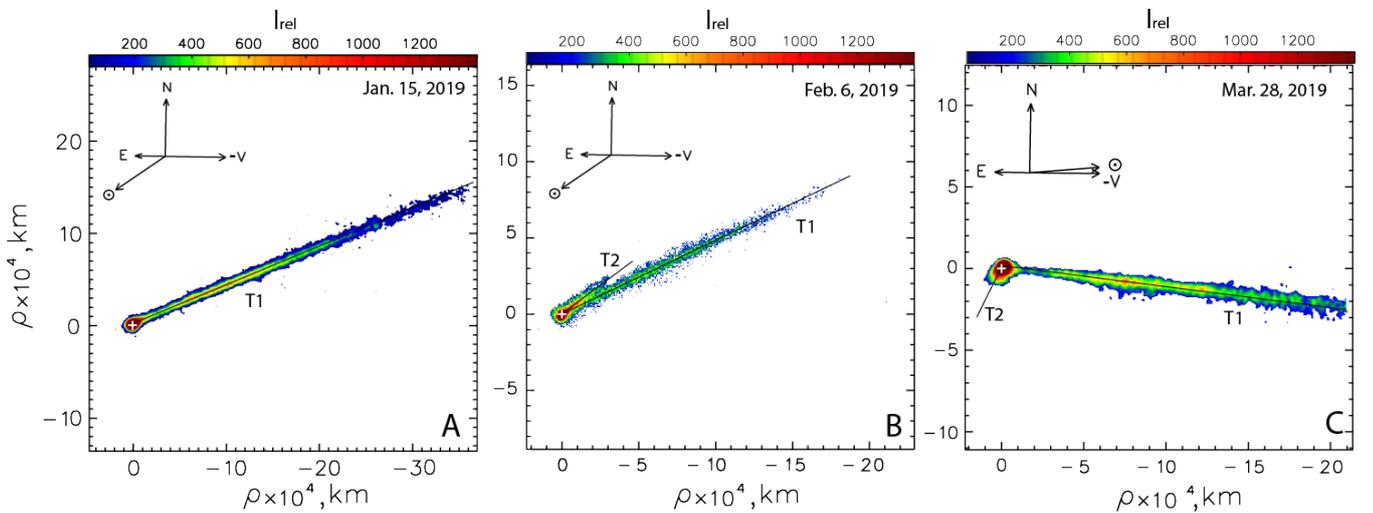

**Fig. 2.** The composite images of asteroid (6478) Gault obtained in the R filter. Panel (A) shows the image taken at the 2.5-m telescope (CMO) of SAI MSU on January 15, 2019; panels (B) and (C) show the images taken at 1.3-m telescope (SPb) on February 6, and March 28, 2019, respectively. The optocentre of the asteroid is marked by a white cross. Black lines show the directions of the tails (T1 and T2).

The T1 directions are $PA$=291.5°, 294.8° and 264.3° for observations made on January 15, February 6, and March 28, 2019, respectively. The T2 directions are $PA$=306.8° and 165.7° for observations made on February 6 and on March 28, 2019, respectively. The obtained directions are close to the results presented by Jewitt et al. 2019 and Kleyna et al. 2019.

The orientations of the tails correspond to the geometry obtained from the comet toolbox[1] (Vincent, J.-B., 2014). This toolbox is based on the Finson and Probstein theory (Finson and Probstein 1968a,b), where only particles released in the orbital plane of the body with zero initial velocity are considered. However, it provides a good approximation of the tail shape. For the illustration, we present in Fig.3 a comparison of synchrones and syndynes built for the summarized images of the asteroid obtained on February 6, 2019.

To compare our observations with the already published results for Gault, we estimated the grain sizes using equation (Kleyna et al. 2019).

$$\beta = 5.740 \times 10^{-4} \times \frac{Q_{pr}}{\rho \cdot a} \qquad (1)$$

where $Q_{pr}$ is a radiation-pressure efficiency coefficient (∼1–2 for rocky and icy material), $\rho$ is the density, and $a$ is the grain size.

As one can see from Fig. 3, the dust emission producing the tails started 102 days before our observation on February 6, 2019, for tail T1 and 38 days for tail T2, correspondingly. Our values are in good agreement with the results presented by Kleyna et al. 2019 and Ye et al. 2019. Tail T1 is longer than tail T2 (see Fig.2) and consists of particles with sizes from 20 to 200 μm.

---

[1] http://www.comet-toolbox.com/FP.html

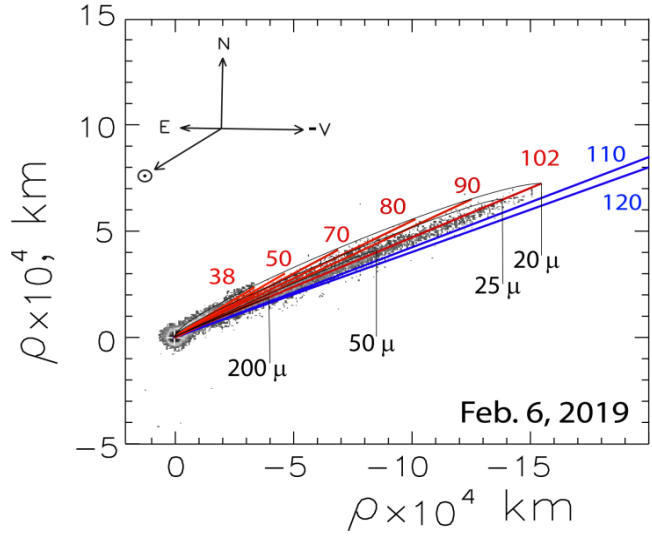

**Fig. 3**. The image of the (6478) Gault (obtained on February 6, 2019) and the composite Finson-Probstein diagram, constructed via the comet toolbox (Vincent, J.-B., 2014). Syndynes (black) and synchrones (red+blue), are labeled in days the observation. Arrows point to the direction to the Sun (☉), North (N), East (E), and the negative velocity vectors (-V).

*Magnitude, Lightcurve, Colour*

For the analysis of the dust environment of the asteroid Gault, we used broadband filters. The integrated magnitude $m_a$ can be evaluated from the equation

$$m_a = -2.51 \lg\left[\frac{I_c(\lambda)}{I_s(\lambda)}\right] + m_s - 2.51\ p(\lambda)\Delta M, \qquad (2)$$

where $\rho$ is the aperture of radius, $I_a$ and $I_s$ are the measured fluxes of the asteroid and a standard star in counts, respectively, $m_s$ is the star magnitude, $p(\lambda)$ is the sky transparency depending on the wavelength, $\Delta M$ is the difference between the air masses of the observed asteroid and star. As we applied field stars for calibration, the sky transparency is not considered.

Table 3 Photometry, reduced magnitudes, and colours of the comets.

| Date, UT | r [au] | $m_R^*$ | B-V | V-R | B-R | $Af\rho^*_R$ [cm] | Instrument |
|---|---|---|---|---|---|---|---|
| 2019 Jan. 15 | 2.456 | 17.42±0.03 | 0.77±0.04 | 0.40±0.04 | 1.17±0.04 | 47±1 | 2.5-m CMO |
| 2019 Jan. 16 | 2.454 | 17.41±0.03 | 0.79±0.04 | 0.40±0.04 | 1.19±0.03 | 47±2 | 2.5-m CMO |
| 2019 Jan. 18 | 2.450 | 17.41±0.02 | 0.80±0.04 | 0.42±0.03 | 1.22±0.04 | 46±1 | 2.5-m CMO |
| 2019 Feb. 06 | 2.411 | 17.53±0.03 | - | 0.42±0.03 | - | 34±1 | 1.3-m Pleso |
| 2019 Feb. 08 | 2.407 | 17.41±0.03 | 0.70±0.10 | 0.53±0.10 | 1.22±0.04 | 38±1 | 1.3-m Pleso |
| 2019 Mar. 23 | 2.316 | 17.08±0.05 | - | 0.49±0.04 | - | 39±2 | 0.61-m Pleso |
| 2019 Mar. 27 | 2.307 | 17.22±0.04 | - | 0.42±0.06 | - | 35±1 | 1.3-m Pleso |
| 2019 Mar. 28 | 2.305 | 17.32±0.02 | - | - | - | 32±2 | 1.3-m Pleso |
| 2019 Jan. 10 | 2.466 | **17.42±0.01 | 0.78±0.03 | 0.40±0.03 | 1.18±0.03 | 48±1 | Jewitt et al.,2019 |
| 2019 Feb. 21 | 2.382 | **17.20±0.01 | 0.75±0.03 | 0.41±0.03 | 1.16±0.03 | 42±1 | Jewitt et al.,2019 |
| Sun | - | -27.15±0.01 | 0.63±0.02 | 0.39±0.01 | 1.02±0.02 | - | Willmer, 2018. |

* for calculation we used aperture radius ~ 5000 km

** in (Jewitt et al., 2019) aperture radii 5717 km and 4457 km were used for January 10 and February 21, respectively.

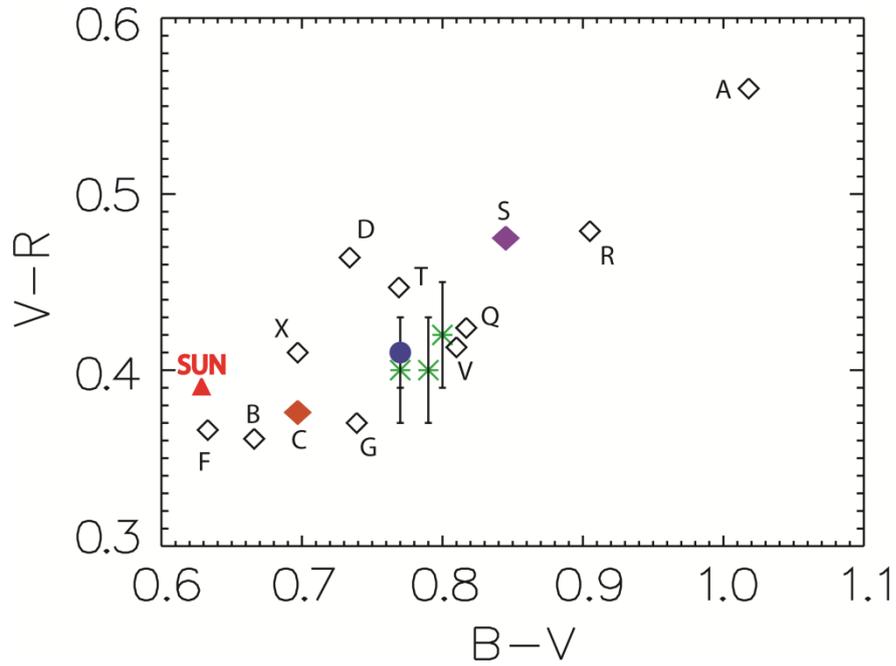

**Fig. 4.** Colour-colour diagram of asteroid Gault. Our results are marked by green asterisks with error bars. The result from Jewitt et al. (2019) is marked by the blue circle. The red triangle shows the colour of the Sun from Willmer (2018). We also show results (marked by diamonds) for asteroids of various spectral types A, B, C, D, F, G, Q, R, S, T, V (data from Dandy et al., 2003).

Using the direct images in the broad-band filters B, V and R one can evaluate the dust colour. The resulting colour-colour BV-VR diagram is shown in Fig. 4.

Gault usually refers to a Phocaea family and is considered as a stony S-type asteroid. However, Jewitt et al. (2019) concluded that in an optical range this asteroid is more similar to C-type asteroids. The retrieved broadband optical and near-infrared colours and optical spectroscopy (Bolin, 2019) also suggest that Gault belongs to the C-complex asteroids. From our results (Fig. 3) we can see that the colour of the asteroid is bluer than that of S-type and redder than that of C-type asteroids. Recently, Marsset et al. (2019) confirmed that Gault is a silicate-rich (Q- or S-type) object likely linked to the Phocaea collisional family. This asteroid exhibits substantial spectral variability over the 0.75–2.45 μm wavelength range (Marsset et al. 2019): from unusual blue (S′ = −13.5 ± 1.1% μm$^{-1}$) to typical red (S′ = +9.1 ± 1.2% μm$^{-1}$) spectral slope. These colour variation does not seem to correlate with activity.

Based on the observations, we built a colour map and analysed colour variations along the tail for the observation made on January 15, 2019. Each pixel of the summed image was converted into the apparent magnitude and the resulting BV, VR and BR colour maps were constructed by subtracting the two sets of images from each other. An average error in the magnitude measurements was 0.04$^m$. The resulting colour maps of the Gault asteroid are shown in Fig. 5 (panels a,b,c). These maps allowed us to make a cross-section along the tail to see colour variations. One can see (Fig. 5, panels d,e,f) that the innermost near-asteroid region has a red colour. The colour profiles show a colour gradient: the colour of the coma decreases sharply within a range of ~10 000 km for all filters, and after this, the colour value is increasing at a distance of ~20 000 km. Beyond 25 000 km, the colour practically is not changed on average showing random scattering only. This behaviour of colour is of great interest and we will examine it in detail in Section 6.

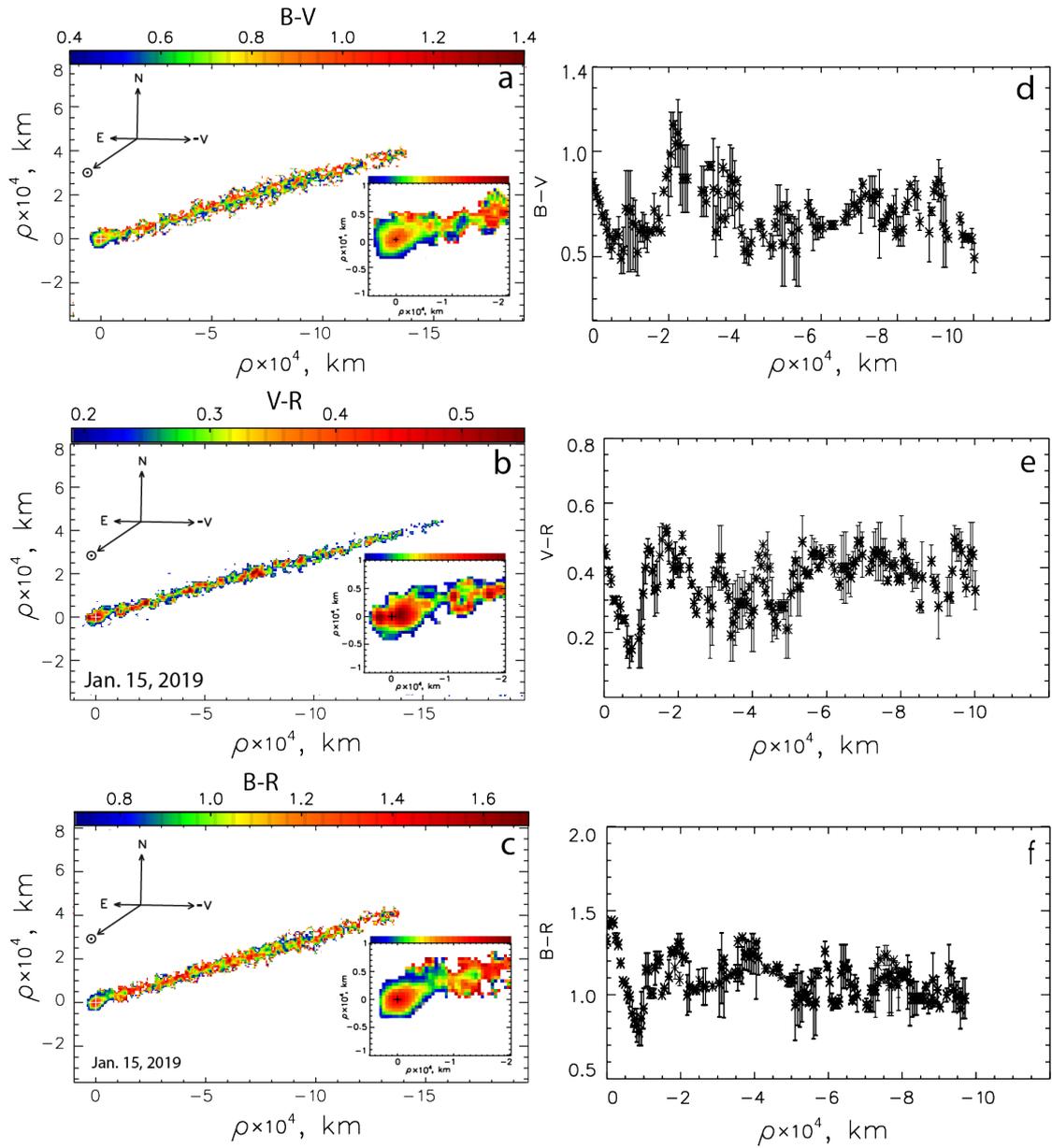

**Fig. 5.** BV (A), VR (B) and BR (C) colour maps of the asteroid Gault constructed for the images obtained on January 15, 2019 are shown in the left column (panels a,b,c). The maps are coloured according to colour indices in magnitudes, as indicated in the colour bar at the top of the image. The projected directions toward the Sun, North, East, and the negative velocity vector are indicated. The optocentre is marked by a cross. The corresponding radial profiles of the dust colours along the tail in the different filters are shown in the right column (panels d,e,f). The error in the colour measurements varies from 0.02 to 0.1$^m$. The central parts of the coma are shown in the insets.

Because dust activity is a common feature for asteroid Gault and comets, it seems attractive to use an approach that is widely applied in the physics of comets to estimate the degree of their activity. The so-called parameter *Af ρ* introduced by A'Hearn et al. (1984) is generally used to characterize the dust abundance in the coma via a measure of the solar radiation reflected by the dust. The value *Af ρ* [cm] (where *A* is the albedo of the dust grains, $f$ is the filling factor of the grains within the field of view, and $\rho$ is the radius of the aperture at the active asteroid Gault) can be used to estimate the dust production, assuming a steady-state dust outflow model. In this case, the radial surface brightness is expected to decrease with $\rho^{-1}$ and the $Af\rho$ value should be aperture independent. However, it is known that the surface brightness variation deviates from the reciprocal of the projected distance in the plane of the sky, and the $Af\rho$ value, therefore, varies with distance from the optocentre (Jewitt & Meech, 1987). We calculated $Af\rho$ according to equation (A'Hearn et al., 1984):

$$Af\rho = f \frac{4r^2 \Delta^2 10^{0.4(m_{Sun} - m_a)}}{\rho}, \qquad (3)$$

where $m_{Sun}$ and $m_a$ are the Sun and asteroid magnitudes in the wavelength band of observations, respectively (see Tab. 3), $r$ is the heliocentric distance in [au], $\Delta$ is the geocentric distance in [cm], $\rho$ is the radius of the aperture in [cm]. The $Af\rho$ values were calculated for the R filter within a circular aperture of about 5000 km radius. The corresponding results are shown in Table 3. Our results vary from 47 to 32 cm for the period from January to the end of March 2019, respectively. For comparison, the value $Af\rho$ for Gault presented in the CARA data for the period from February 5 to April 17, 2019 change from 47 to 10 cm for an aperture radius of 4000 km, respectively. The estimated $Af\rho$ values for Gault are more similar to the results obtained for the active Jupiter Family Comets at heliocentric distances from 2 to 3 au (Lowry et al., 1999) than for new long-period comets (Mazzotta Epifani et al. (2006, 2010, 2011, 2014), Meech et al. (2009); Szabó et al. (2002; 2008), Ivanova et al., 2015, 2016).

In Fig. 6, we present the results obtained on January 15, 2019 for the radial profile and the derived $Af\rho$ values for the three different filters, respectively. We show the values obtained for the entire flux, as it is observed. Clearly, there is a strong dependence on the radial distance. Perhaps the $Af\rho$ change would be less significant if the nucleus contribution was subtracted. However, we believe that the real $Af\rho$ values are more important than the ones estimated after subtraction of the nucleus, because there is a contribution from the coma even at the maximum of brightness. A similar trend with a decrease of $Af\rho$ is observed for some distant comets at large heliocentric distances (Korsun et al., 2014; Ivanova et al., 2019), where this feature is probably explained by the fragmentation of the emitted dust particles. But the fragmentation mechanism is most probably not applicable in the general case of asteroids.

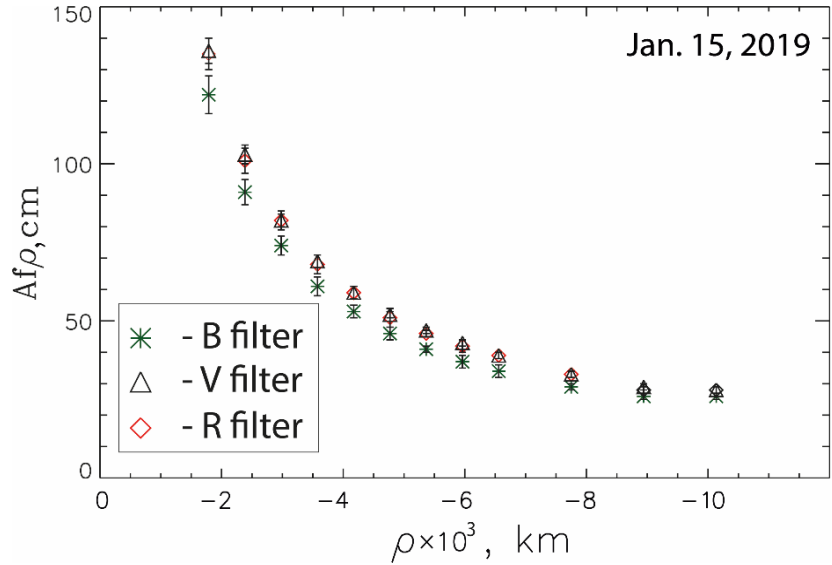

**Fig 6.** Dependence of the *Afρ* values on the radial distance, derived from co-added images obtained on January 15, 2019, with the 2.5-m telescope of CMO.

Certainly, in case with asteroid Gault parameter *Afρ* cannot be associated directly with dust production, as it used for comets. The parameters must correlate with dust production only in the unlikely case that the size distribution and the ejection velocity are time-independent (see e.g. Fulle, 2000). However, we believe that evaluation of this parameter is useful: first, for the quantitative estimation of the dust activity and its comparison with the corresponding activity of comets at similar heliocentric distances; second, our evaluations can be compared with previously published values of this parameter (Ferrín et al., 2019) that allows us to see variation activity.

*Period of rotation of the asteroid*

On February 06 and March 28, 2019, two longer photometric runs (based observations at Skalnaté Pleso Observatory only) were performed at the Skalnaté Pleso Observatory (see Table 3) to analyze the temporal behavior of the light curve and to determine the rotational period. Only the R filter was used for this purpose. Our observations, lasting more than 2.5 hours, show an almost constant light curve without any noticeable periodic variations of brightness (Fig. 7). The results of observations made on the other night are also shown in the plot to emphasize the flat type of the light curve. The flat shape of the composite light curve can be related to a nearly spherical shape of the asteroid or small variations of the projection area seen from the Sun and Earth. A contribution of light from the emitted dust grains masking the surface also may play a role. Also, one cannot exclude in advance that the rotation period is much longer than the observation time.

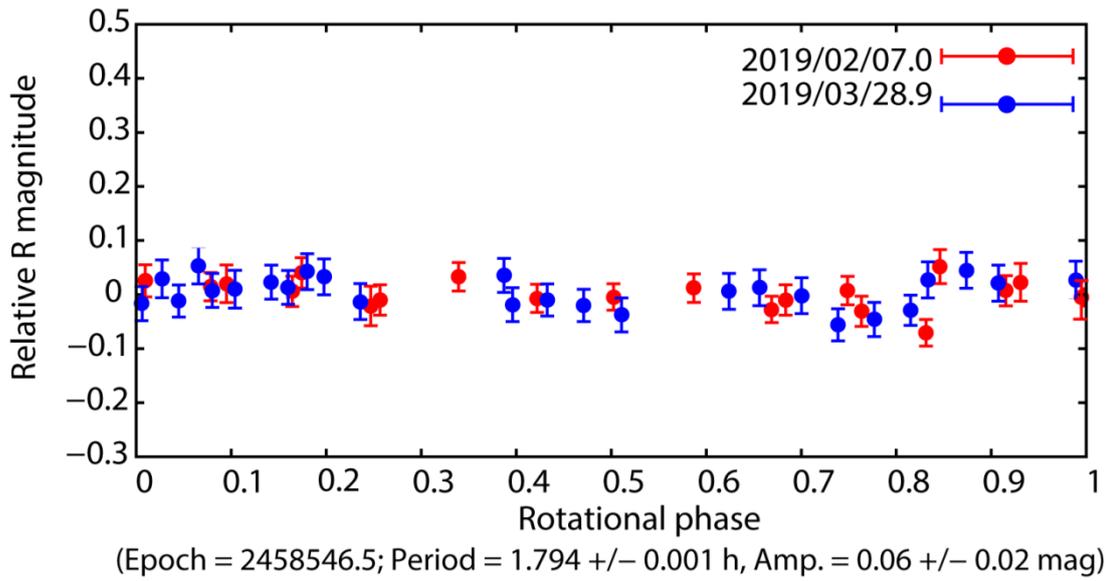

**Fig. 7.** Composite lightcurve of the active asteroid (6478) Gault (based observations at Skalnaté Pleso Observatory). Representative magnitude error bars (±0.019 mag) are shown. The mid-time of the observational runs are shown in the legend.

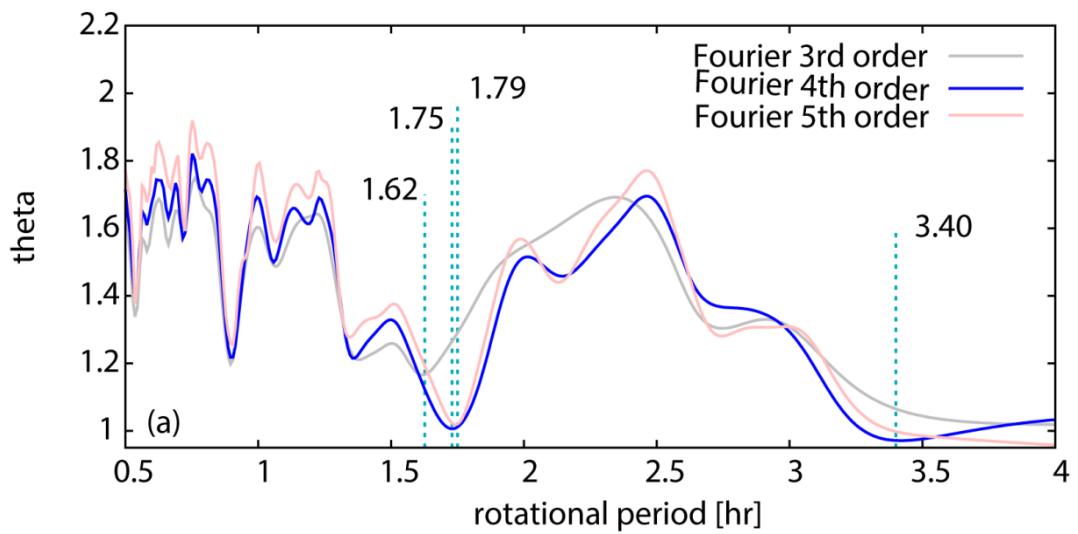

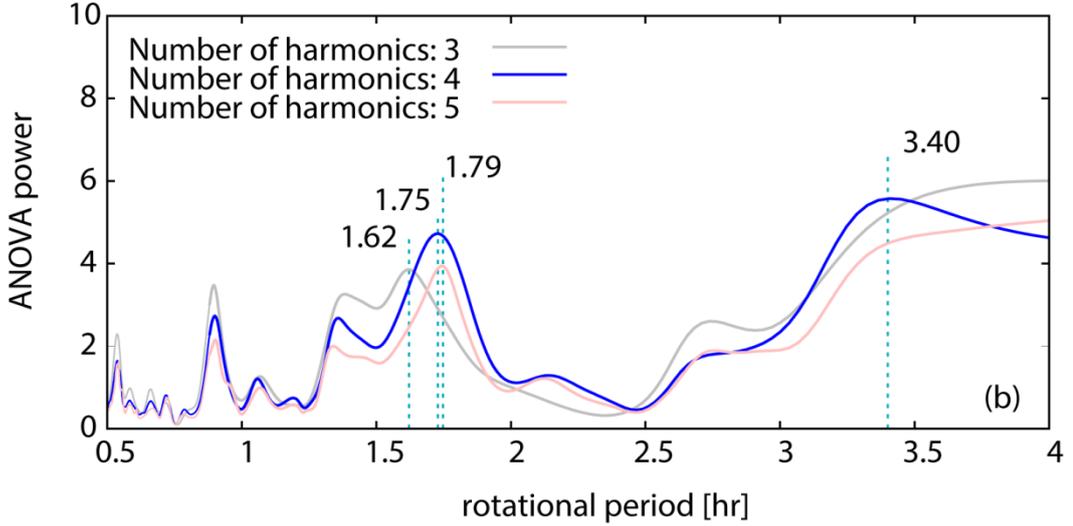

**Fig. 8.** FALC (a) and ANOVA (b) periodograms (scan of periods from 1 to 4 hours) applied for phased lightcurve data from two nights. The theta value represents the root-mean-square dispersion of the observed data and approximated Fourier series, the ANOVA power maxima show probable values of the rotational period.

To identify the periodicity in the observations, we use the technique described by Harris et al. (1989) and known as the FALC method. In addition to this approach, we analyzed the photometric data using a four-Fourier-component Analysis of Variance (ANOVA) (Schwarzenberg-Czerny, 1996). The first probable value (see Fig.8) of the rotational period (1.79 hr from FALC, 1.79 hr from ANOVA) is very close to the value of a ~2-hour rotational period confirmed by Kleyna et al. (2019). Another probable period is close to 3.4 hr and it could confirm the analysis of Ferrín et al. (2019) or Carbognani and Buzzoni (2020). Moreno et al. (2019) analyzed five light curves showing a similar flat behavior of brightness and concluded that the rotation period couldn't be clearly determined from their observations. The same conclusion was drawn by Jewitt et al. (2019). Note that, Ferrín et al. (2019), analyzing the observed light curve, made another surprising conclusion that Gault could be a binary asteroid. If it is so, Gault is the first active asteroid with its own satellite. Of course, such circumstance greatly complicates the analysis of the brightness curve and should be thoroughly checked. We believe that the accurate value of the rotation period cannot be determined at the moment. The next good apparition for a detailed photometric study of Gault will open from August to October 2020. It could reach a maximum brightness of 16.8 mag in opposition before the end of September 2020.

*Estimation of the asteroid diameter*

The observations allow to derive an approximate value of the asteroid size based on standard procedures. The absolute magnitude $H_V$ can be evaluated from the apparent magnitude $m_V$ according to the formula

$$H_V = m_V - 5\log(r\Delta) + 2.5\log(\Phi(\alpha)) \qquad (4),$$

where $r$ and $\Delta$ are the heliocentric and geocentric distances in au and $\Phi(\alpha)$ is the phase function at phase angle α. We use the classical expression $\Phi(\alpha) = (1 - G)\Phi_1(\alpha) + G\Phi_2(\alpha)$, where $G = 0.15$. For the calculation of the effective diameter, the formula from Bowell et al. (1989) is applied

$$\log(p_V) = 6.259 - 2\log D - 0.4 lH_V \quad (5),$$

where $D$ is the diameter of the asteroid in kilometers, $p_V$ is the geometric albedo, and $H_V$ is the V-band magnitude. Because the accurate albedo of Gault is unknown, a value of $p_V = 0.22$ is adopted, which is a common estimation for the Phocaea family (Carruba 2009; Nesvorný et al. 2014). Using this approach, we find that a upper limit of diameter of the asteroid (6478) Gault is 3.9±0.1 km. Jewitt et al. (2019), assuming that *2.5log(Φ(α))= -0.04α*, found a value that is about 0.2 km smaller than our result.

## 4. Evolution of the orbit of asteroid Gault in the main-belt

At the beginning of 2019, about twenty active asteroids were known. During the first quarter of 2019, five new members were added to this list. Gault is one of them. As it was mentioned above, the activity of the main-belt asteroids may be caused by different processes. One of the possible explanations of Gault's activity is the gravitational capture of a comet into the main-belt. As a "former comet", this asteroid could conserve a noticeable amount of volatiles, causing its repetitive activity. To investigate this scenario, we integrated Gault's orbit 100 kyr into the past.

The parameters of the orbit of asteroid Gault was taken from the JPL Small-Body Database Browser (Giorgini et al., 1996) and used for the starting orbit in our integration and all further integrations. The heliocentric ecliptic orbital elements of Gault referred to the equinox J2000.0 are $q = 1.85890$ au, $a = 2.30515$ au, $e = 0.19359$, $\Omega = 183.5576°$, $\omega = 83.26769°$, $i = 22.811322°$, for epoch 2458600.5.

To clarify Gault's evolution within the main-belt, we tested several possible evolutionary scenarios that could be implemented with different degrees of confidence. In addition to Gault's nominal orbit, we considered the evolution of 100 clones of its orbit, representing the statistical uncertainty. For this purpose, we used the method developed by Chernitsov et al. (1998). We took the orbital elements of the nominal orbit in the form of a covariant 6 × 1 matrix $y_o$. Then, the corresponding covariant matrix with the elements of the orbit of the j-th clone, $y_j$, can be calculated as $y_j = y_o + A\eta^T$. Here, A is a triangle matrix such that the product $AA^T$ equals to the covariance matrix related to the process of the nominal-orbit determination. The covariance matrix can be found on the website of the JPL browser. Detailed information about this method can be found in Tomko & Neslusan (2019). The numerical integration of the orbits was performed by using the integrator RA15 (Everhart, 1985) within the MERCURY package (Chambers, 1999). The gravitational perturbations of the eight planets (from Mercury to Neptune) are considered. Potential non-gravitational effects are ignored in the integration. The evolution of Gault's orbit and its clones backward in time for 100 kyr are shown in Fig. 9. Particularly, we present the evolution of the perihelion distance (Fig. 9a), the eccentricity (Fig. 9b), and the inclination (Fig. 9c). A similar simulation was performed for the longitude of the ascending node and the argument of perihelion, and the results are completely consistent with our conclusions. The solid red curve shows the evolution of the orbital elements of Gault and the green curves show the evolution of the clones. To simplify the comparison, we artificially shift vertically the red curve for the nominal orbit of Gault in all figures.

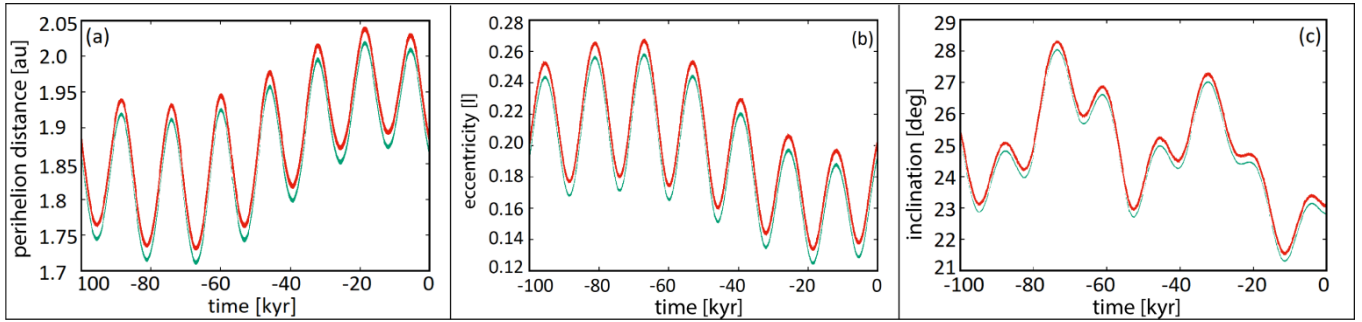

**Fig. 9.** The evolution of the perihelion distance (a), the eccentricity (b), and the inclination to the ecliptic plane (c) of the orbit of (6478) Gault (red curve) and the orbits of the set of its clones (green curves) for up to 100 kyr in the past. The curve for the nominal orbit of Gault (red curve) is artificially shifted over the original position to being visible.

The orbital evolution of all clones shows the same behavior as the evolution of the nominal orbit of Gault. We do not see any visible deviations of the orbital elements of the clones from the elements of the nominal orbit. Based on this, we believe that Gault's orbit has not experienced significant changes in at least the past 100 thousand years. This analysis makes the assumption of the cometary origin of this object unlikely.

Haghighipour et al. (2002) analysed the orbits of the three asteroids, 7968 Elst-Pizzaro, 118401, and P/2005 U1, and a large number of hypothetical mail-belt comets and showed that 7968 Elst-Pizzaro and 118401 are stable for 1 Gyr, whereas the third object, P/2005 U1 becomes unstable in ~20 Myr. Instability of the last object is the result of its orbital proximity to the region of 2:1 MMR with Jupiter, suggests its formation near this resonance. These results can be considered as possible scenarios for the formation and origin of maim-belt comets, including Gault. Our analysis of the orbital evolution of Gault during the last 100 kyr is not in contradiction to their results, because the simulation period (100 kyr) is short for the approaching to the time of the formation for this object. Note that, a long-term orbital evolution is not the focus of our analysis.

## 5  Close approaches of the periodic comets

The idea about the collision between asteroids as a reason for the short-term activity has long been known and generally recognized. In the case of Gault's activity, a rare feature is the repeatability of the observed events (see Fig. 1 and discussions there). We propose to consider active comets as an alternative source of projectiles that could collide with the object in question. To test this hypothesis, we analyzed the orbits of numerous comet candidates and evaluated the possibility of collisions with cometary debris (meteoroid streams originating from these comets). This research is based on the numerical integration of the cometary orbits and the orbit of Gault and tracking their mutual positions. The orbits of 380 short-periodic comets (labeled from 1P/Halley to 380P/PANSTARRS) with the orbital elements published by JPL[2] are tested. As was suggested in Neslušan et al. (2016), we are looking for the cases when the distance between the orbits of a candidate and Gault is within 0.15 au, and the corresponding heliocentric distance of the comet is small enough for a meteoroid stream to be produced. As a result, material can be released from the asteroid, leading to the observed dust-tail formation. To find potential candidates, we numerically integrated Gault's orbit as well as the orbits of all tested periodic comets up to 1000 years into the future. As before we used the integrator RA15 and the gravitational perturbations of the eight planets, Mercury to Neptune, are included. During the tested period, 102 periodic comets can approach Gault's orbit at least once.

According to Chandler et al. (2019) and Hui et al. (2019), previous cometary features were detected in 2013 Sep. 22., 2013 Sep. 28., 2013 Oct. 13., 2016 June 9., 2016 June 10., 2017 Oct. 23., 2017 Nov. 11., 2017 Nov. 12., and 2019 Jan. 5. Based on these dates we looked for a potential rapprochement between the orbit of a candidate and Gault's orbit in 10 day-intervals before and after the observed activity, because the beginning of Gault's activity was not fixed accurately. Given these constraints, four periods of activity, namely: + (2013 Sep. 12 - Oct. 23), (2016 May 31 - June 20), (2017 Oct. 13 - Nov. 23) and (2018 Dec. 26 - 2019 Jan. 15) were examined, assuming that the corresponding activity manifestations can be potentially explained by the impacts.

As a result of the analysis, we found 41 periodic comets approaching Gault's orbital arc at least once in the above four time intervals interesting for us. Among these candidate comets, 36 comets were found in all intervals, i.e. the condition of the orbits approaching within 0.15 au is fulfilled for all dates. For these comets, the minimum approach distance varies from 0.07635au (73P/Schwassmann-Wachmann 3-Y) to 0.14933au (13P/Olbers). The case of the comet 73P is truly intriguing: calculations showed that a "hypothetical collision" between the test objects took place sometime after the actual fragmentation of 73P firstly confirmed by Boehnhardt and Kaufl (1995). The minimum approach distance between Gault and the orbit of comet 73P/Schwassmann-Wachmann 3-Y is about 0.07635 au for the four considering time intervals. Very interesting results were obtained for the main-belt object/comet 311P/PANSTARRS showing occasional activity in 2013. The images taken from the Hubble Space Telescope in 2013 September revealed a set of 6 linear, comet-like tails (Jewitt et al., 2018). Its activity was interpreted as a result of an equator-ward landslide from the surface of the rotating asteroid (Jewitt et al., 2013, 2015b; Hirabayashi et al., 2015; Hainaut et al., 2014). Our study shows that the orbit of 311P is very close to Gault's orbit. Because the activity of 311P was observed during about 250 days in 2013 (Jewitt et al., 2018), the collisions of meteoroids ejected from the nucleus of 311P with Gault cannot be ruled out. The evolution of the minimum distance between the orbital arcs is shown In Fig. 10. One can see a quasi-periodic decreasing and increasing of the minimum value caused by a different orbit profile. It is worth noting that the proximity of the orbital periods of these objects becomes important.

---

[2] https://ssd.jpl.nasa.gov/?sb_elem

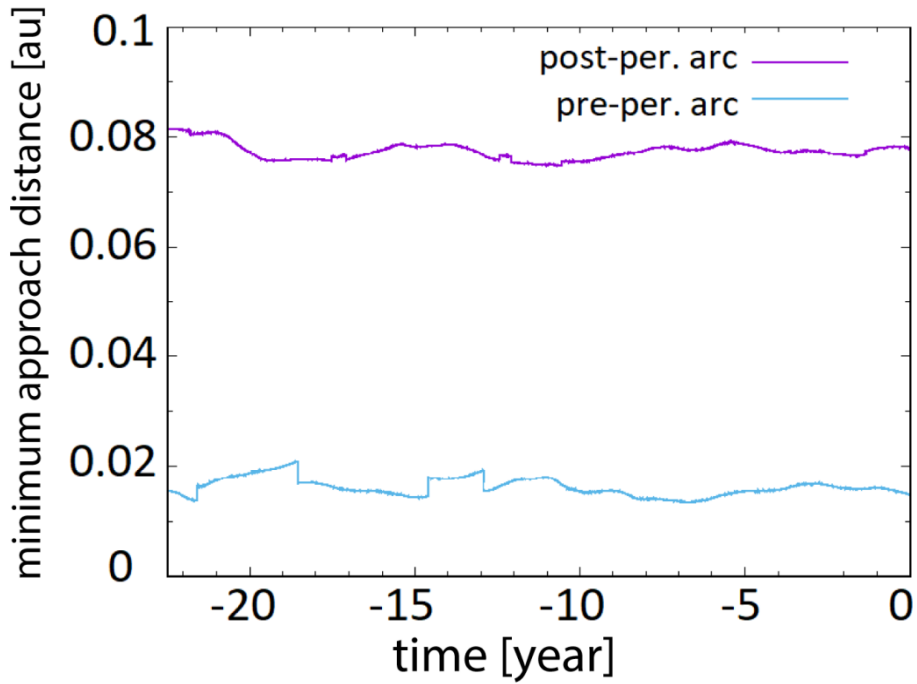

**Fig. 10.** Evolution of the minimum distance between the orbital arcs of Gault and comet 311P/PANSTARRS from 22.5 years before the present up to the present. The minimum distance of the post-perihelion (pre-perihelion) arc is shown by the violet (cyan) curve.

## 6. Discussion and Conclusions

Observations of active asteroids have specific features. Unlike regular observations of comets, which exhibit long-term and usually slowly varying gas/dust activities, observations of active asteroids often occur for a limited period due to observation conditions and the relatively short duration of the phenomenon. This leads to the natural overlapping of observations. In our opinion, this does not diminish the value of independent observations, especially when the original tools are used, as is the case in our study. In this paper, we presented the results of observations of active asteroid (6478) Gault, performed in the optical range with telescopes ranging from 0.61-m to 2.5-m in imaging mode between January 15, and March 28, 2019. The dates of specific observations complement a set of already published observations (e.g., Chandler et al., 2019; Ferrín et al., 2019; Jewitt et al., 2019 Hui et al., 2019; Lee et al., 2019; Marsset et al., 2019; Moreno et al., 2019; Sanchez et al., 2019; Ye at l., 2019; Kleyna et al., 2019). Given this, we first briefly dwell on our results, which confirm published earlier ones.
Two distinct dust comet-like tails were detected during the observations. The morphology of (6478) Gault is typical for most of the active asteroids (133P, 238P, 311P): it is a point-like nucleus with a long thin dust tail. A similar morphology was observed in the previous active phase of the asteroid in 2013 and 2016 (Chandler et al., 2019). The analysis of particle dynamics confirms the conclusion published by Moreno et al. (2019): the dust emissions occurred approximately in November 5, 2018 and January 2, 2019 and had a short duration (about 5 days). The observed structures are well described by a model with relatively large particles (from a dozen microns to sub-millimeter). An interesting question is the distribution of particle sizes. We will return to it below. Our estimate of the size of the asteroid also agrees well with that obtained by Jewitt et al. (2019). Somewhat more unexpected is the result of evaluating the rotation period. This characteristic attracted a lot of attention since one of the hypothetical mechanisms explaining the activity is associated with the fast rotation of the object. It has been suggested (Kleyna et al. 2019) that low cohesion and rapid rotation can cause surface material redistribution and dust emission. As noted earlier (Moreno et al. 2019), the light curve is very flat, not allowing to confidently assume the rotation period. Such behavior can be explained, for example, by a slightly varying projection area. This assumption is not related to the real shape of the object: As the calculations performed for comet 67P (Marshall, 2019) show, bodies with a very bizarre shape can have an almost constant projection area seen from the sun. In addition, the absence of significant brightness variations can be associated with a slow rotation, when the rotation period is much longer than the observation time. Using special tools, we got estimates that are consistent with the estimates given by Kleyna et al. (2019). Nevertheless, these results should be treated with caution, because, for example, the application of the Fourier analysis to a time-limited observation will inevitably show the presence of some maxima. It should be noted the original hypothesis put forward by Ye et al. (2019), where the authors suggest a binary system for the asteroid Gault. We believe that testing this hypothesis and a more reliable estimate of the rotation period requires new extended observations.

Now we turn to the analysis of new results. Undoubtedly, observations of colour behavior are of great interest. As known, for example, from cometary observations (Zubko et al 2015), colour slopes provide information both on the composition and on the distribution of particle sizes (especially when it comes to observing the coma). Our observations performed in wide comet filters showed that the asteroid is visibly redder than the sun (see Fig. 3). As noted above, although Gault traditionally is referred to a Phocaea family and is considered as a stony S-type asteroid, its optical properties have the features of C-type asteroids, which was noted by Jewitt et al. (2019) and Bolin (2019). This suggests that carbonaceous and silicates are present in its composition. Taking into account all the data presented in the figure, we conclude that Gault does not clearly exhibit properties that make it possible to unambiguously attribute it to a certain type of asteroids. No less intriguing are the colour maps shown in Figure 4. First of all, the pronounced red colour of the dust coma near the asteroid attracts attention. Note that this redness can be associated both with the lack of small particles (such an assumption was made by Moreno et al. (2019)), and with a change in the composition of the dust particles scattering the radiation. For comets, the excess of red is well manifested if the exponent in the particle size distribution function is approximately equal to 2 (see Fig. 5 in Zubko et al. 2015). This interpretation is in good agreement with a detailed analysis of the dynamics of the tails. So the exponent is equal to: –2.28 (for grains between 1 μm and 15 μm) and –3.95 (for grains between 15 μm and 870 μm) in Moreno et al. (2019); -2.5 to -3 (for grains between 10 μm and 20 μm) and -4 for larger particles (Ye et al 2019). Note that the calculations show a very weak dependence of colour on the composition for exponents higher than 3. It is these large values (~ 4) that are assumed for large particles in Moreno et al. (2019).

In addition to the red colour of the inner region, a quick change in colour as the distance from the object increases is of great interest. The colour indices BV, VR and BR fall down significantly with increasing distance from the asteroid in the region inside ~10 000 km. This change is clearly visible both in colour images and in the sections shown in Figure 4. Note that this change has a pronounced trend, relative scattering is small. Similar changes in colour have been repeatedly observed in cometary comae (e.g., Korsun et al. 2016). As mentioned above, this effect can be associated both with a change in the particle size distribution (for micron particles that scattering efficiently in the optical range, an increase in the fraction of small particles can cause the coma to turn blue) and with the release of particles having different composition (and therefore another refractive index). For the asteroid, the second scenario seems less likely. A rapid relative increase in the number of small particles for comets is often associated with fragmentation of particles as they move away from the cometary nucleus. However, the fragmentation mechanism usually proposed for comets (namely, the evaporation of ice fractions in dust particles) is obviously not applicable in the general case of asteroids. The only option is when activity occurs due to sublimation of ice near the surface. We will show below why this option is not suitable for Gault. Finally, it can be assumed that the colour change reflects size stratification arising due to the difference in the velocities of dust particles: small dust particles have a noticeably higher speed and their deficiency is formed near the asteroid and the relative excess at large distances. A more rigorous verification of this assumption requires specifying the mechanism of dust particles ejection, which will make it possible to quantify their velocities. Such a study is beyond the scope of this paper.

Our attempt to use the $Af\rho$ characteristic to describe the activity of asteroid Gault is of interest. This parameter is widely used in cometary studies, primarily for the quantitative comparison of activity. Low measured values of $Af\rho$ (from 32 to 46 cm) indicate weak activity in comparison with typical dust production for comets at a similar heliocentric distance (Lowry et al., 1999; 2003). We should remember that originally the $Af\rho$ concept was introduced for cometary coma in steady state (A'Hearn et al., 1984), when $Af\rho$ is an aperture-independent parameter. For our case, when the central part of the images (r $<10^4$ km) is quasi-spherical and quasi-homogeneous (see boxes in Fig. 4), its use seems reasonable. Contrary to expectations, we received a noticeable drop with distance from the optical centre (Fig. 5). This decrease is observed in all filters. The simplest analysis suggests that this behavior can be explained by several scenarios. For example, reducing the effective area of the diffusers will lead to this effect. The decrease in area can be related in turn both to a strong decrease in the particle size (if the size of the scatterers becomes noticeably smaller than the wavelength of the incident radiation), and to the formation of large clusters, which (while maintaining the total dust mass) have a smaller area and scatter radiation less efficiently. We believe that both of these options do not work for the asteroid. As a simple alternative, a noticeable acceleration of particles in the region under consideration can be proposed. In this case, the invariance of the number of scatterers in the circular segment is not satisfied, namely this condition is crucial for the invariance of $Af\rho$. But even this assumption seems doubtful. Even for comets, in which there is a near-nuclear zone of acceleration of dust by the outflowing gas, the size of this region is usually a few tens of the sizes of the nucleus, i.e. much less than a thousand kilometers. It remains to be assumed that the observed noticeable decrease in $Af\rho$ indicates a strong real heterogeneity of the internal region, despite the photometric uniformity, which may simply be the result of projection. Additional studies are required to understand whether it is possible to obtain certain restrictions on the morphology of activity from a joint study of the photometric fields of colour and *Afρ*. We will address this issue in future work.

Undoubtedly, one of the most intriguing questions is the question of the cause (or causes) that initiated the observed dust activity. Below we will try to consider the main mechanisms discussed in the asteroid community. Let us first dwell on the mechanism when the cause of dust release is heating of the surface of the asteroid and sublimation of volatile components. First of all, we note that the obtained spectra (Lee, 2019) did not show the presence of emissions of water and carbon dioxide. However, it gives us only an upper limit for a possible gas production rate. The above colour analysis will not allow excluding the existence of volatiles. Another compelling argument against the hypothesis of an icy cause of activity is our analysis of the evolution of the orbit of Gault. We showed that with a high probability this object is located inside the snow line for a very long time. This suggests that even if ice had been present in its evolutionary composition, the surface region should be deprived of it. Finally, note that the orbit of Gault has a slight eccentricity. This means that the amount of absorbed energy varies slightly, and it is difficult to allow seasonal activity.

The second possible mechanism causing activity is the loss of material connectivity and its displacement due to rapid rotation. This mechanism was proposed for Gault, for example, by Kleyna et al. (2019). We believe this idea is attractive and fresh. At the same time, we do not have sufficient grounds to back it up. Firstly, to assess the possibility of a loss of connectivity, it is necessary to know not only the size of the body and its rotation period (which determines the maximum centrifugal force) but also the structure of the surface layer and its composition. Laboratory experiments show that even in the case of large (millimeter particles) tensile strength of aggregates is about several pascals (Skorov and Blum, 2012), which is much more than centrifugal acceleration can provide. If we assume that cohesion is completely absent (which, generally speaking, can be allowed for a material deposited on the surface at a very low speed), then the discontinuity of activity cannot be explained. The body, the surface of which is covered by an unbound regolith, must constantly lose substance. These theoretical problems are supplemented by observational difficulties. We believe that the observed brightness curve does not allow a reliable estimate of the period of rotation of the asteroid. New observations are needed.

In the end, we will focus on the mechanism that seems most likely to us. We are talking about multiple collisions colour caused by the approach of an asteroid with meteor showers produced by comet activity. We showed above that there are several candidates whose orbits are close to the orbit of Gault during the periods of observed activity. For some of these candidates (for example, 73P and 380P), the resulting distances are very small. We do not claim that this mechanism successfully explains all the cases of activity presented in Fig. 1, but two clusters seem to be in good agreement with this hypothesis. In Kleyna et al. (2019), Moreno et al., (2019) and Ye et al. (2019), an estimate of the total loss of matter during the observed release was obtained based on an analysis of the morphology of the tails. These results are in good agreement and suggest that the total ejection mass was about $10^7$ kg. This value allows us to estimate the total size of the impactor (or impactors, since in the zeroth approximation we can assume that the mass of emissions adds up). Taking the qualitative estimates for a high-speed impact into a porous target (refs from my reply), we can conclude that the total impactor volume is on the order of a few cubic meters, which seems reasonable (Vogler and Fredenburg, 2019; Sachse et al., 2015). More accurate estimates will be obtained in a future work, where we analyse the relative velocities of objects and perform collision modelling.

.

## ACKNOWLEDGMENTS

OI thanks the DAAD Program, and the Slovak Academy of Sciences (grant Vega 2/0023/18). This article was supported by the realization of the Project ITMS No. 26220120029, based on the supporting operational Research and development program financed by the European Regional Development Fund. DT thanks the Slovak Grant Agency for Science, VEGA, grant No. 2/0037/18, and by the Slovak Research and Development Agency under contract No. APVV-16-0148. IL thanks the SAIA Programme for financial support. The observations at 2.5-m telescope of CMO were conducted using the equipment acquired through the funds of the Program for Development of the Moscow State University. The authors thank the anonymous reviewer for valuable comments on the manuscript.